# Making sense of Open Data Statistics with Information from Wikipedia

Daniel Hienert[1], Dennis Wegener[1], Siegfried Schomisch[1]

[1] GESIS – Leibniz Institute for the Social Sciences,
Unter Sachsenhausen 6-8, 50667 Cologne, Germany
{Daniel.Hienert, Dennis.Wegener, Siegfried.Schomisch}@gesis.org

**Abstract.** Today, more and more open data statistics are published by governments, statistical offices and organizations like the United Nations, The World Bank or Eurostat. This data is freely available and can be consumed by end users in interactive visualizations. However, additional information is needed to enable laymen to interpret these statistics in order to make sense of the raw data. In this paper, we present an approach to combine open data statistics with historical events. In a user interface we have integrated interactive visualizations of open data statistics with a timeline of thematically appropriate historical events from Wikipedia. This can help users to explore statistical data in several views and to get related events for certain trends in the timeline. Events include links to Wikipedia articles, where details can be found and the search process can be continued. We have conducted a user study to evaluate if users can use the interface intuitively, if relations between trends in statistics and historical events can be found and if users like this approach for their exploration process.

## 1 Introduction

Nowadays, a mass of open data statistics is available from providers like Eurostat, the United Nations or The World Bank. Eurostat for example offers about 4,500 different statistics with a wide topical range from General and Regional Statistics, Economics and Finance, Population and Social Conditions, Industry, Trade and Services, Agriculture, Forestry and Fisheries, International Trade, Transportation, Environment and Energy or Science and Technology.

Visualization software on the Web, such as Gapminder [22], makes use of these statistics and presents them in interactive graphics like diagrams or maps. Users, from laymen to politicians, can explore the data and find trends and correlations in order to "easily improve their understanding about the complex society"[1].

However, statistics - even interactively visualized - are not self-explanatory. They reflect trends for a statistical indicator (*what*) in a certain time period (*when*) and for various countries (*where*). But, these trends often base on or are related to certain events in the topic of the statistic. From a user's perspective, one can ask *why* there is a certain trend. Which influential factors lead to the trend shown in the statistic or may be related? Additional information, topically related to statistics, can help users to interpret them, get background information and hints where to continue in the search process.

---
[1] http://www.gapminder.org/faq_frequently_asked_questions/

We have created a prototype user interface that gives access to thousands of open data statistics from The World Bank, Eurostat and Gapminder. Users can interactively explore these statistics in different diagrams and map views. Additionally, for the Gapminder statistics, we aim at finding topically related historical events to present them in a timeline.

However, there is not much meta-information about the statistic except the title or sometimes a category. Using only keywords from the title leads to very few results when querying related information like historical events. Therefore, we use a query expansion method based on the Wikipedia and the DBpedia corpus to expand a search query with additional related terms based on the statistic title. Statistical visualizations and a timeline with the related events found are then interactively connected in a user interface to let the user explore relationships and explanations for trends in the statistic. Groups of events found with query expansion can be hidden or revealed with facets. In a user test we evaluate if users are able to use the prototype for the exploration of statistical data and related information.

Section 2 gives an overview of related work. Section 3 presents our approach for the combination of open data statistics and topically related historical events. Section 4 presents a user study to examine the exploratory search process. We discuss the results and conclude in Section 5.

## 2 Related Work

A lot of different data providers give access to open data statistics with focus on the world development, e.g. the United Nations, Eurostat or The World Bank. Most datasets contain information on a statistical indicator (often with several dimensions) for countries over time. Gapminder for example aggregates data from several providers and offers about 500 different indicators for about 200 countries and with a temporal coverage beginning from 1800 AD till today. Some of these data providers have recently also developed web-based visualization tools for their statistical data. Most of these web applications use maps and line/bar charts to show indicator development over time and for different countries. Gapminder uses a different approach and shows two related indicators in a scatterplot. Similar to statistical software for scientists, such as SPSS, STATA or R, one can see correlations between these indicators. Gapminder additionally uses an animation to show the development of correlation over time for all countries of the world.

For the semantic modeling of events in RDF a number of ontologies is available, e.g. EVENT[2], LODE [24], SEM [8], EventsML[3] and F [23]. A comparison of such ontologies can be found in [24]. Furthermore, there exist ontologies and systems for the annotation of events in the timeline. Gao & Hunter [7] make use of ontologies not only for the semantic markup of events, but also for the modeling of timelines, relationships between events and for the annotation of events of different timelines. Therefore, not only events can be referenced in the Linked Data Cloud, but also relations and annotations. In a web-based interface, users can search for geological events with a high impact such as earthquakes, tsunamis and volcanic eruptions, can analyze relationships and make connections between them with annotations.

---

[2] http://motools.sourceforge.net/event/event.html
[3] http://www.iptc.org/site/News_Exchange_Formats/EventsML-G2/

The simple representation of events in timelines can be implemented with various web-based tools such as SIMILE[4] widgets. The use of timelines can cause user interaction difficulties, for example, if too many events are visualized and the user loses the navigational overview [16]. Kumar et al. [15] propose a theoretical model for timelines on the storage and presentation layer with levels like content, operations, browsing and editing. An example implementation of this model allows the creation, visualization and browsing of bibliographic metadata. LifeLines [21] utilizes the arrangement of several timelines in one view. Facets of personal records, either flat or hierarchic, can individually be selected by buttons or trees to give an overview or discover relationships. Other complex timeline visualizations like Semtime [14] or Sematime [26] display in a similar fashion several stacked timelines with additionally depicted time-dependent semantic relations between events. Advanced interaction techniques include hierarchical filtering and navigation like zooming or expanding sub-timelines. Sense.us [9] is a web-based visualization system with an emphasis on the social and collaborative aspect. Users can share visualizations, e.g. a chart of US census data, and can discuss and comment on different states of the visualization. Bookmarking ability of different states and graphical annotation tools allow annotating and discussing certain data points or trends in statistical visualizations in order to make sense of the pure data. ChronoViz [6] is a system that uses the timeline metaphor for the simultaneous display of multiple data sources such as video, audio, logs, sensor data and transcriptions. Also here, users can add annotations with a digital pen and use them as anchor links. Aigner et al. [2] give a systematic overview of visualizing time-oriented data based on categorization criteria on the time, data and representation level. They emphasize open problems and future work like multiple views of time-oriented data and their coordination. In this sense, some case studies exist which use Multiple Coordinated Views [28] of temporal and other visualizations in domains like climate data [25] or medicine [1]. Timelines are linked to other views, so that selecting data in a timeline highlights data in other views or vice versa.

The search for relationships between pieces of information in different representations can be described by the model of Exploratory Search [17]. Multiple iterations with cognitive processing and interpretation of objects over various media like graphs, maps, text and video is needed. The user has to spend time "scanning/viewing, comparing and making qualitative judgments" for the result of "knowledge acquisition, comprehension of concepts or skills, interpretation of ideas, and comparisons or aggregation of data and concepts" [17]. Also, timelines can be part of an exploratory search process, for example, to use the time component to explore research articles in a timeline view [3].

Query expansion is used to expand a search query with additional terms like synonyms, related terms, or methods like stemming or spelling correction. For very short queries this can increase the overall recall of a search, and more relevant information objects can be found [27]. Query expansion based on thesauri, ontologies, co-occurrence analysis or other knowledge sources have been utilized in digital libraries of different domains like e.g. the social sciences [18] or the medical domain [13]. Also, the Wikipedia corpus has been used as a database for this purpose [19].

---

[4] http://www.simile-widgets.org/timeline/

## 3   System Prototype

We have created a research prototype[5] for the visualization of open data statistics and related events. In the following, we will describe (1) the visualization of statistical data in several views, (2) the retrieval and visualization of related historical events.

### 3.1   Visualization of Statistical Data

The web application is implemented in PHP and the visualization and interaction component is realized with HTML5, the canvas element and JavaScript (based on a previous prototype presented in [12]). We have integrated statistical data from Eurostat, The World Bank and Gapminder. The Eurostat dataset contains 4,545 indicators, The World Bank 3,566 and Gapminder 498.

Because the prototype already contains about 8,500 statistics from different providers, it is difficult to provide an overall hierarchical categorization. Therefore, the web application provides a query interface for searching statistics by title. The user can enter keywords and an autocomplete function initially suggests matching statistics. By clicking on the link the indicator is chosen and instantly visualized.

Similar to existing solutions, the statistical data is displayed in various graphical views: a map, a bar chart and a line chart (see Figure 1).
On a Google map, indicator values for all different countries and a given year are visualized as circles of different sizes and colors at the country location. The values of the indicators are visually encoded in two dimensions: (1) the higher the indicator value, the greater the circle radius, (2) the higher the indicator value, the more the color shade goes into the red spectrum, low indicators values are encoded with bluish tones. This way, the user can instantly see where high values are clustering. For example, for the indicator *Fertility*, one can see at first glance that in 2011 still more children per woman were born in Africa than in Europe. The map has standard interaction facilities like zooming and panning, so the user can browse to the region of interest.

The bar chart shows indicator values for each country and a given year in descending order. Countries with high values are at the top, countries with low values are at the bottom of the list. As 200 countries are listed, the user can scroll the list. The width of the horizontal bars allows seeing differences between countries instantly, so the users can compare if there are only small or big differences between two countries in the list. Similar to the map, the bars are color-coded, so one can see instantly if countries have higher or lower values than the Median.

---

[5] http://opendatastatistics.org

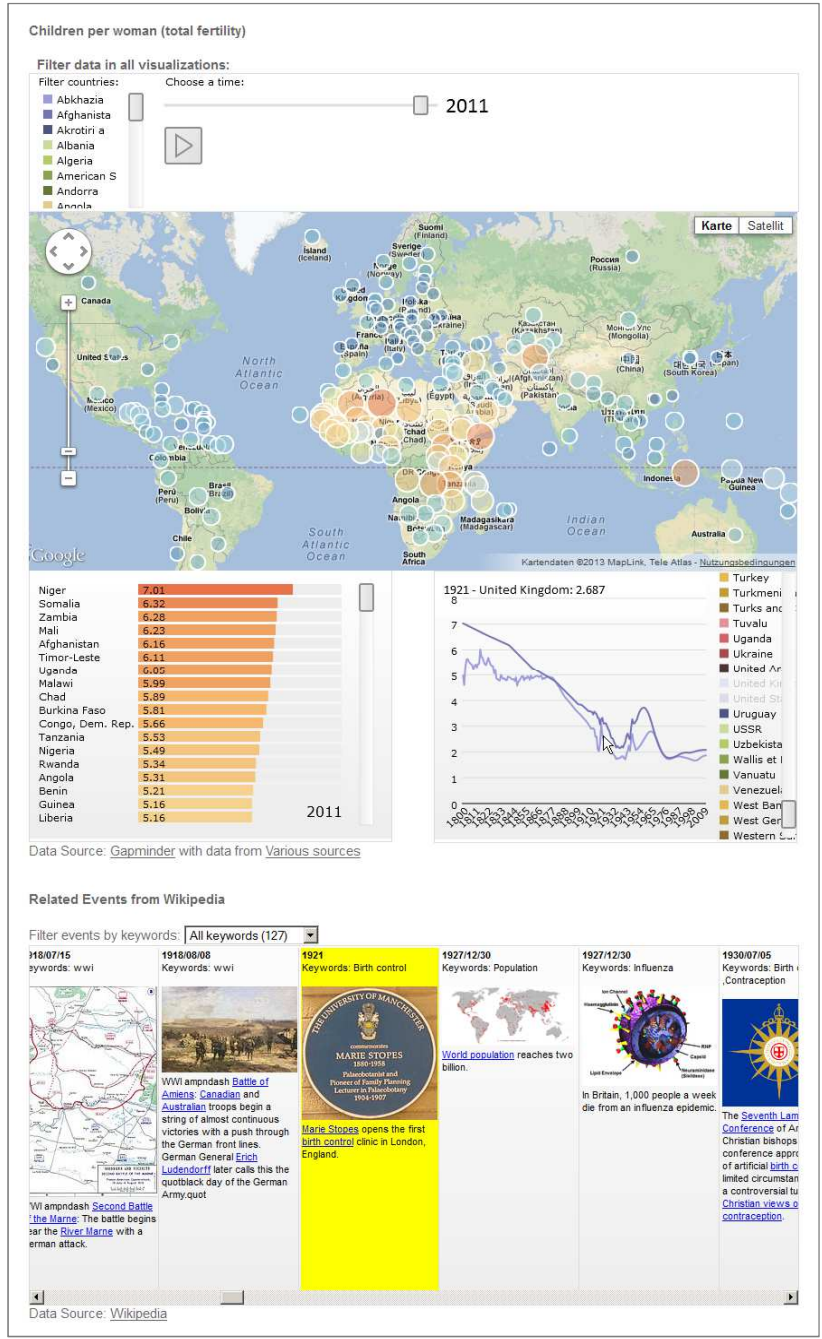

**Fig. 1.** Prototype with statistical visualizations and a timeline with related events for the statistic "Children per woman (Total fertility)" from Gapminder. Hovering the mouse over the data point United Kingdom – 1921 scrolls to and highlights events for the same year in the timeline.

Next to the bar chart, there is a line chart that shows the distribution of indicator values over time. Users can select a country from the chart legend or from the overall filters which results in the line showing up in the graph. Hovering with the mouse over a data point shows time, country and value at the top of the chart. Lines are color-coded similarly in the legend and in the graph.

At the top of the page, there exists a control panel for filtering all views simultaneously by one or several countries and a year. From a list, users can select individual countries which are color-coded according to countries in the line chart. In addition, there exists a slider for selecting the year. This way, the selected year for the map and bar chart view can be chosen and the user can browse through the time. With a play button all views can be animated to show data for the actual year.

### 3.2 Retrieval of Related Historical Events

A goal in this paper is to retrieve and display historical events that are related to a statistical indicator and therefore, link the following two datasets:

1. The Gapminder dataset includes 498 different statistical indicators. This dataset is maintained regularly, topically well chosen, includes many data sources and covers all world countries with a wide temporal coverage from 1800 till today.
2. The historical events dataset is based on a database that holds a large collection of historical events extracted from Wikipedia [10] and can be queried via an API[6]. It has been shown that machine learning with features from DBpedia is a feasible way to achieve an automatic classification of the extracted events into categories [11]. The outcome of the effort was a dataset with about 190,000 historical events covering different languages and including category information. The data subset used in this paper contains 37,859 yearly English events from 300 BC to 2013. An important requirement is that both datasets have a focus on years as a temporal reference unit. Statistical indicators are resolved for years; historical events are chosen by the Wikipedia community as an important fact for a year's history.

The statistic title or its category only provide weak information for the retrieval of related events. For example, querying the historical database for the statistic "Children per woman (total fertility)" with the keyword "fertility" leads to 0 results. The query has to be expanded with strongly related concepts to find more events and to get a higher recall. For "fertility", concepts like "contraception", "birth control", or "syphilis" must be found to retrieve historical events that have an explanatory character for trends in the statistic. The two data sources are connected in the following way (see Figure 2):

*Step 1*: For each of the Gapminder statistics, the title is preprocessed by removing units and stop words/characters.

---

[6] http://www.vizgr.org/historical-events/

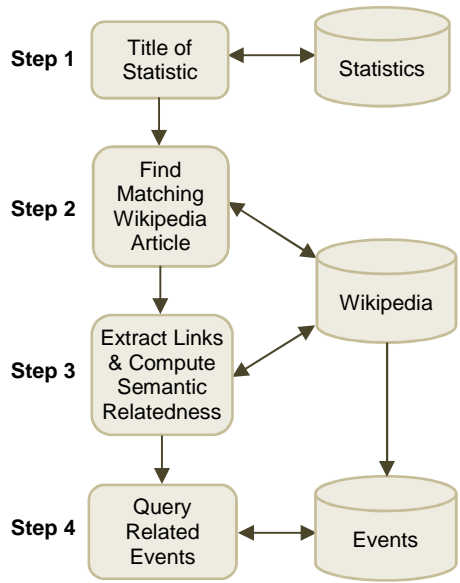

**Fig. 2.** Process steps to retrieve related events for a statistic.

*Step 2*: Based on the preprocessed title, the Wikipedia API is queried for an article page with a matching concept. We inspected the top 10 search results and manually selected the semantically best fitting page for each statistic. As a result, we got a mapping to a chosen Wikipedia article for each statistical indicator. All in all, this includes 144 distinct articles, as some of the statistics belong to the same Wikipedia concept (e.g. to the Wikipedia article "Development aid"). Because statistic titles from Gapminder are really short, we decided to manually select the Wikipedia article to guarantee a well-chosen concept with balanced properties of abstractness level, size, number of links etc.

*Step 3*: As a next step, we want to find related concepts for the mapped concepts. Therefore, we have implemented a web service[7] that returns related concepts for an input term. The service queries in/outlinks from Wikipedia via the Web API as well as broader/narrower terms and category information from DBpedia [4] via SPARQL endpoint. For each concept, the semantic relatedness (SR) to the original term is computed. For doing so, we use the Normalized Google Distance (NGD) formula [5], but instead of taking hit counts from a search engine we use hit counts from the Wikipedia full text search.

---

[7] http://www.vizgr.org/relatedterms/

Semantic relatedness is then computed with the following formula:

$$SR = \frac{log_{10}\big(\max(A,B)\big) - log_{10}(A \cup B)}{log_{10}(W) - log_{10}\big(\min(A,B)\big)}$$

$A$: Number of full text search hits in Wikipedia for concept one
$B$: Number of full text search hits in Wikipedia for concept two
$A \cup B$: Number of full text search hits in Wikipedia for concept one AND concept two.
$W$: Number of articles in Wikipedia.

In a separate evaluation with evaluation datasets included in the sim-eval framework [20] we found that this approach achieved a Spearman correlation up to 0.729 for human judged datasets and P(20) up to 0.934 for semantic relation datasets. We use all concepts with a SR > 0.3 as keywords for the following step, which filters very broad or non-related concepts. This way, we could compute 77,202 related concepts for 498 statistics with on average 155 keywords per statistic.

*Step 4*: In the last step, we query events based on the concepts found in the previous step. In detail, we query the Historical Events-API for all events that include at least one of these keywords in the event description and lies within the time interval of the statistic. For all 498 statistics, querying only the Wikipedia equivalent for a statistic returns in sum 8,450 events, which means on average about 17 events per statistic. Using the presented approach for query expansion, for all statistics in sum 137,921 (not distinct) events are returned, which means on average about 279 per statistic. Based on the returned events a timeline is build. To illustrate the different steps, we present two examples related to the topics *fertility* and *earthquakes*.

Example *"Fertility"*:
The title of the statistic from Gapminder in the first example is "Children per woman (total fertility)". The title was mapped to the Wikipedia concept "Fertility". Querying related concepts returns 61 keywords with an SR value higher than 0.3 (compare Table 1). These keywords were then used to query the Historical Events-API which results in 22 matching events.

Example *"Earthquake"*:
The title of the statistical indicator in the second example is "Earthquake - affected annual number". For this statistic, the Wikipedia article "Earthquake" was selected. 132 related concepts were returned with an SR value higher than 0.3. For these keywords the Historical Events-API returned 1,008 events.

### 3.3 Visualizing Related Historical Events

Below the area that visualizes statistical data, a scrollable timeline shows queried historical events in a timeline (see Figure 1, at the bottom). The individual events consist of a date, keywords, a description and, if available, a thumbnail. Links in the event description allow browsing to Wikipedia articles and to read details there.

Table 1. Top 20 Wikipedia concepts including values for semantic relatedness for "fertility" and "earthquake".

| Links for Fertility | SR | Links for Earthquake | SR |
|---|---|---|---|
| Fertility rate | 0.714 | Tsunami | 0.693 |
| Total fertility rate | 0.657 | Seismic | 0.651 |
| Infertility | 0.638 | 2011 Tohoku earthquake and tsunami | 0.584 |
| Sperm | 0.622 | Seismology | 0.581 |
| Crude birth rate | 0.590 | Epicenter | 0.572 |
| Contraception | 0.588 | Aftershock | 0.566 |
| Fertile | 0.577 | Disaster | 0.554 |
| Fertilization | 0.576 | 2004 Indian Ocean earthquake | 0.549 |
| Gestation | 0.572 | 2010 Haiti earthquake | 0.543 |
| Uterus | 0.571 | 1906 San Francisco earthquake | 0.529 |
| Fertility clinic | 0.561 | Northridge earthquake | 0.523 |
| Ovary | 0.559 | Volcano | 0.521 |
| Menstrual cycle | 0.558 | Megathrust earthquake | 0.518 |
| Menopause | 0.552 | Subduction | 0.512 |
| Pregnancy | 0.544 | Seismologist | 0.510 |
| Endocrinology | 0.535 | Natural disaster | 0.507 |
| Fetus | 0.535 | Landslide | 0.502 |
| Fecundity | 0.533 | Foreshock | 0.496 |
| IUD | 0.513 | Mercalli intensity earthquake | 0.491 |
| Sperm count | 0.512 | 1755 Lisbon earthquake | 0.489 |

Events for a certain keyword can be filtered with a select box above the timeline. The line chart that shows the indicator distribution over time for certain countries and the timeline are linked by a brushing-and-linking mechanism. When a user brushes over a data point in the line chart, the timeline automatically scrolls to events from the same year and highlights them with a yellow background or vice versa.

## 4  User Study

We have conducted a user study to examine the following research questions:
- Can participants use the interface intuitively?
- How do users normally search for additional information of trends in a statistical indicator?
- Can our approach of showing related historical events in a timeline help users to find interesting background information and starting points for further exploration?
- Which advantages and limitations does our approach have?

### 4.1 Method & Participants

The participants were asked to carry out a set of tasks in the prototype and to fill out a questionnaire after each task with their actions performed, results found, time required, experienced difficulty level and comments. First, the users could familiarize themselves with the environment for 5 minutes. After that, they had to complete three different tasks. Finally, the users were asked to evaluate the pros and cons of our approach and to assess the overall scenario. The group of participants included eight male researchers and one female researcher, aged 27 to 40 (mean: 31 years). All had a graduate degree in computer science like Master or similar. The participants were asked to rate their experience in dealing with web-based search on a five-point-scale. The rating was 1.56 ("very good") with a standard deviation of 0.73.

### 4.2 Task & Questions

The participants had to handle the following tasks and answer questions in the questionnaire:

*1. Fertility trends across countries*
Briefly describe the development of fertility rates in the United States of America and in the United Kingdom from 1921 to 1940 as shown in the line graph (a). What are the values for each country in the years 1926 (b) and 1937 (c)?

For task 1, the users had to write down the response in a text field, answer how long the process took, assess as how difficult the task was perceived on a five-point-scale (2=very easy, 1=easy, 0=normal, -1=difficult, -2=very difficult) and were asked to give comments and suggestions.

*2. Causes for the decrease in fertility after 1920 and after 1960 in the US and in the UK*
(a) Find possible reasons for the decrease of fertility in the United States of America and in the United Kingdom from 1920 to 1940. (b) Find possible reasons for the decrease of fertility in the United States of America and in the United Kingdom in 1960. Try to find information outside the given prototype using other sources, e.g. Google. Do not spend more than 5 minutes for each subtask (a) and (b).

For task 2, the users had to record the search steps, the total time, the relevant information sources and a confidence score for the search result on a five-point-scale (1=very unsure, 2=unsure, 3=normal, 4=sure, 5=very sure). Furthermore, similar to task 1, they had to assess the difficulty and could give comments and suggestions. For this task, the timeline has not been visible in the UI to prevent any influences on the user. For task 3, the participants could then refresh the user interface and activate the timeline with a GET-Parameter in the URL.

*3. Usage of the timeline*
Analog to task 2, please find potential causes or contexts for the decrease in fertility rates in these countries from 1920 to 1940 and from 1960 based on the historical events displayed in the timeline and describe them briefly.

For task 3, users had to write down the response in a text field, enter a confidence score as in task 2 and to assess the difficulty and could give comments and suggestions as in tasks 1 and 2.

*4. Comparison and evaluation of the two methods from tasks 2 and 3 and overall results*

Please evaluate the overall scenario:
- What are the pros and cons of both search methods for the application scenario?
- Which search strategy would you favor for the given application scenario and why would you do so?
- Was the integrated user interface including the graphical view of information of fertility statistics and historical events in a timeline helpful for answering the questions on the decrease of fertility after 1921/1960 in both countries?

The overall scenario could be rated with a five-point-scale (2=very helpful, 1=helpful, 0=normal, -1=not helpful, -2=not helpful at all) and we left room for general comments, suggestions and criticisms.

### 4.3 Results

After the participants had made themselves familiar with the user interface they could all (n=9) solve task 1 successfully. The average time exposure for sub task (a) resulted in 74 seconds, while answers for the country specific values (b) and (c) took on average 16 seconds (compare Table 2). Participants rated the difficulty level of the task on average with "easy" (1.00). Task 1 showed that the user interface could be easily adopted by users for filtering and read-off processes without further explanation. Nevertheless, some points of criticism and improvement were given in terms of better scrolling functionality in the country list, more precise handling of the sliders, better highlighting of the selected countries, more distinguishable colors and zooming functionality in the line chart.

For task 2, the participants recorded 30 query steps in the questionnaire to find possible reasons for the decrease of fertility in the US/UK between 1920 and 1940 (a) and from 1960 (b). The majority of users used Google (23 times), followed by Wikipedia (6 times). Solving task 2 took 122s/118s per identified reason. Users stated Wikipedia (10 times) and other websites (18 times) as sources of information. On average, the participants had a normal confidence in the information found on the Web (3.19/3.29). The average difficulty level of task 2 was evaluated with "difficult" for (a) and "normal" for (b).

In task 3 the participants were faced with the same questions as in task 2, but now could enable and use the timeline in the user interface to find possible reasons. For the decrease in fertility from 1920 to 1940 in the USA/UK, participants named, e.g., nine times the event "*1921 - Marie Stopes opens the first birth control clinic in London, England.*", five times *World War I and II* (from several events) or two times "*1930/07/05 - The Seventh Lambeth Conference of Anglican Christian bishops opens. This conference approved the use of artificial birth control in limited circumstances, marking a controversial turning point in Christian views on contraception.*" One user stated that a decrease in fertility is normal for industrial countries. For task a) the participants were confident for possible reasons ("sure", 3.67).

**Table 2.** Summarized Results

| Task 1: Fertility trends across countries | | | |
|---|---|---|---|
| | Successful responses | Time needed | Difficulty level |
| a) Trend from 1921 to 1941 | 9/9 | 74s | „easy" (1.00) |
| b)/c) Values for 1926 and 1937 | 9/9 | 16s | |

| Task 2: Causes for the decrease in fertility after 1920 and after 1960 in the US and in the UK | | | | | |
|---|---|---|---|---|---|
| | Query steps | Relevant sources | Confidence | Time needed per reason | Difficulty level |
| a) From 1920 to 1940 | 12x Google; 3x Wikipedia; 1x general knowledge | 12x different sources; 4x Wikipedia | "normal" (3.19) | 122s | "difficult" (-0.50) |
| b) From 1960 | 11x Google; 3x Wikipedia | 8x different sources; 6x Wikipedia | "normal" (3.29) | 118s | "normal" (-0.25) |

| Task 3: Usage of the timeline | | | | |
|---|---|---|---|---|
| | Potential reasons | Confidence | Time needed per reason | Difficulty level |
| a) From 1920 to 1940 | 9x: Marie Stopes opended first birth control clinic in UK in 1921; 5x: World War I/II; 2x 7th Lambeth Conference; 1x: normal development for industrialized countries | "sure" (3.67) | 58s | "normal" (0.39) |
| b) From 1960 | 8x: FDA - Approval and usage of oral contraceptive pill; 2x: baby boom ends; 1x: Vietnam war; | "sure" (4.17) | | |

For the decreasing trend from 1960 the users stated, e.g., eight times "*1960/05/09 - The U.S. Food and Drug Administration announces that it will approve birth control as an additional indication for Searle's Enovid, making it the world's first approved oral contraceptive pill.*", two times "*1958/12/31 - Based on birth rates (per 1,000 population), the post-war baby boom ends in the United States as an 11-year decline in the birth rate begins (the longest on record in that country).*" or one time *Vietnam War* (from one event). For task b) the participants were confident in having found possible reasons (4.17, "sure"). The task needed 58 seconds per reason and was perceived as "normal" (0.39) difficult on average.

For task 4, users could first compare "search on the Web" (task 2) and "search in the prototype" (task 3). The web search was advantageous to the users in terms of using familiar systems (Google, Wikipedia etc.), having access to all kinds of documents, sources and contexts, the possibility to compare the resulting information and the control of the search process. The disadvantages reported were the heavy-handed search methods to find the relevant documents or sources, the time consuming process, additional search steps, no single answers for complex questions, a lot of irrelevant information and the problem of how to translate the information need into a search query when no starting point is given.

For the search in the prototype, the participants stressed as pros that they have found the results faster and could see them directly. Furthermore, they assessed positively the single point of search as well as the integrated search environment and the synchronization between line chart and timeline. In contrast, they qualified negatively that the search quality in this system depends on the used data sources and that only Wikipedia documents were offered but no scientific studies or other information.

Seven participants would initially prefer the prototype from task 3 to get an overview over a topic and to get results quickly. Some participants stated that this information was enough; most participants explained that they would use the prototype for a quick overview and then use the web in order to get reliable and quotable data or for verification of the results of task 3.

The participants evaluated the search in the prototype tendentially as helpful, which became apparent in the average value of 0.50 ("helpful"). This was also stressed, although with constraints, in the textual evaluation. One participant thought the prototype to be convenient only for easy questions but not for complex issues in the social sciences, other users wished more data sources.

## 5   Discussion & Conclusion

The conducted user study gives first hints that the combination of numerical statistical data in different views and related information like events in a timeline in a single user interface can be fruitful. Participants were able to use the interface intuitively without further instructions. The results of task 2 show that participants search for related information of trends in statistics with search engines like Google and Wikipedia with a combination of what, when, and where keywords. It can be seen from the confidence values that the users were confident about having found related information. On the Web, they have access to all kinds of documents, but it was a time-consuming process and a lot of search steps were needed. Our prototype was preferred for a fast overview as a single point of search and a starting point, but, of course, did not include all sources a web search engine provides.

For the user test, we chose the very complex topic "fertility" in contrast to less complex topics like e.g. "earthquakes" (in a sense of how many other diverse complex aspects may have an influence on the indicator). Influential factors are difficult to determine for a normal web search user, because related concepts and keywords are not at hand. Table 1 gives an overview: While concepts for "fertility" are very diverse, a lot of concepts for "earthquake" include the term "earthquake" with combinations of different locations and dates. This makes it easy also for normal web search, where the query instantly leads to the matching Wikipedia article (e.g. for "2010 Haiti earthquake"). In contrast, querying for all aspects of fertility seems to be a harder process. If one is lucky, search queries like "fertility 1920s United Kingdom" (as performed by participants of our study) lead to documents were information and explanations for one location and one time period are included. However, this documents most times only provide very broad explanations for a concept and no further detailed information or links. For example, for fertility users found a document that contains information on the development of fertility in the UK for the last century. Here, influential factors like both world wars, economic depression (late 1920s) and influenza outbreak (after World War I) are described. However, the document did not include any further information or links. Instead, users have to copy and paste keywords into a search engine. Our approach tries to take off the burden of searching relevant keywords and querying related information. In this sense, we computed related concepts and used these to query the data source.

In the presented use case, we concentrated on the combination of statistics with historical events. Since these events are carefully chosen according to their importance on a world history view by Wikipedia users, they offer a good compromise between importance, abstractness, count and temporal coverage (compare [10]). Of course, other similar data sources could be included such as news articles from the Guardian Open Platform[8] or the New York Times API[9]. However, we found that these sources provide the majority of articles for the last three decades and the query with keywords returns a mass of only rather important articles. Here, further aggregation steps have to be applied. But also other related information types can be queried with the computed concepts like Wikipedia articles, web sites, studies and surveys from which the statistics are generated, videos, images etc. Another important aspect is that the search on the Web and the search in the interface are not complementary as proposed in the user study, but users can use links in the timeline to continue with their web search.

At the end, it is always a trade-off between presenting too much or too little information, and carefully choosing important information based on only unsubstantial query information like statistic title (and related concepts), country and time.

---

[8] http://www.guardian.co.uk/open-platform
[9] http://developer.nytimes.com/